# Two Cognitive Transitions Underlying the Capacity for Cultural Evolution

Liane Gabora
Department of Psychology, University of British Columbia
Fipke Centre for Innovative Research, 3247 University Way,
Kelowna, BC Canada, V1V 1V7
liane.gabora@ubc.ca https://people.ok.ubc.ca/lgabora/

Cameron M. Smith
Department of Anthropology, Portland State University Portland, OR, 97207
b5cs@pdx.edu https://www.pdx.edu/anthropology/cameron-smith

## Abstract

This paper proposes that the distinctively human capacity for cumulative, adaptive, open-ended cultural evolution came about through two temporally-distinct cognitive transitions. First, the origin of *Homo*-specific culture over two MYA was made possible by the onset of a finer-grained associative memory that allowed episodes to be encoded in greater detail. This in turn meant more overlap amongst the distributed representations of these episodes, such that they could more readily evoke one another through *self-triggered recall* (STR). STR enabled representational redescription, the chaining of thoughts and actions, and the capacity for a stream of thought. Second, fully cognitive modernity following the appearance of anatomical modernity after 200,000 BP, was made possible by the onset of *contextual focus* (CF): the ability to shift between an explicit convergent mode conducive to logic and refinement of ideas, and an implicit divergent mode conducive to free-association, viewing situations from radically new perspectives, concept combination, analogical thinking, and insight. This paved the way for an integrated, creative internal network of understandings, and behavioral modernity. We discuss feasible neural mechanisms for this two-stage proposal, and outline how STR and CF differ from other proposals. We provide computational evidence for the proposal obtained with an agent-based model of cultural evolution in which agents invent ideas for actions and imitate the fittest of their neighbors' actions. Mean fitness and diversity of actions across the artificial society increased with STR, and even more so with CF, but CF was only effective if STR was already in place. CF was most effective following a change in task, which supports its hypothesized role in escaping mental fixation. The proposal is discussed in the context of transition theory in the life sciences.

# Two Cognitive Transitions Underlying the Capacity for Cultural Evolution

## Introduction

Regardless of the extent to which any particular skill such as tool use, language, or mental state attribution is uniquely human, it would be difficult to argue that any other species remotely approaches the human capacity for the cultural evolution of novelty that is accumulative, adaptive, and open-ended (i.e., with no a priori limit on potential variation). Here, *culture* refers to extrasomatic adaptations—including behavior and technology—that are socially rather than sexually transmitted. This paper synthesizes research from anthropology, psychology, archaeology, and agent-based modeling into a speculative yet coherent account of the cognitive transitions underlying human cultural evolution.

Archaeological evidence refers here to the 'material correlates' or 'precipitates' of behavior. Its interpretation has a long and complex philosophical history. While earlier approaches tended to treat artifacts as indicators of varieties of 'progress', 'conditional cognitive archaeology' approaches (sensu Haidle, 2009 and Wragg-Sykes, 2015) seek to reveal the cognitive conditions responsible for artifacts (and other material precipitates of behavior). We share this more contemporary approach to understanding the 'mind behind the artifact'.

We note that there is to some extent a trade-off in the literature between theories based on historical data, and theories that are cognitively sophisticated. This paper aims to synthesize these approaches, using archaeological data and anthropological research as the point of departure for a proposal that is consistent with contemporary psychology. Note also that our theoretical approach is founded on evolutionary principles, but not those of the evolutionary psychology approach of Tooby and Cosmides (Cosmides & Tooby, 1992; Sell et al, 2009). We take a contemporary conception of evolution that incorporates sociality, individual behavioral variation, agency, and creativity, as opposed to viewing individuals as slaves to fitness equations.

### Evolutionary Transitions

Evolutionary processes have multiple modes and tempos (Stanley 1979; Gould 2002), and often generate unexpected outcomes due to nonlinear interactions between such information levels as genotype, phenotype, environment, and even developmental characteristics (Galis & Metz, 2007). In cognitive evolution, evidence of significant change might well be 'smeared' over time and space for many reasons, including lag between initial appearance and demic diffusion, ambiguities in the archaeological and fossil records, and other factors. We suggest that the theory of evolutionary transitions can provide a useful framework for understanding the cognitive changes culminating in behavioral modernity (BM).

Transitions are common in biological evolution (Szathmary & Maynard Smith, 1995) and transition research unpacks fuzzier terms such 'adaptation due to natural selection', aiming "...to analyze trends of increasing complexity" (Griesmer, 2000), and explain the origins of new varieties of information organization. Szathmary and Maynard-Smith's account of the eight major transitions in the history of life remains widely accepted today (Calcott & Sterelny, 2011), with other transitions continuing to be identified, including the evolution of new sexes (Parker, 2004), and new varieties of ant agriculture (Schultz & Brady, 2008), animal individuality (Godfrey-Smith 2011), metabolism and cell structure (DeLong et al., 2010), technology (Geels, 2000) and hominin socialization (Foley & Gamble 2009). Research on the dynamics (e.g., rates and types) of evolutionary transitions shows that despite their variety they exhibit common features: they are (1) rare, (2) involve new levels of organization of information, (3) followed by diversification, and (4) incomplete (Wilson 2010). Szathmary and Maynard-Smith include the transition from "primate societies to human societies " as part of their "Extended Evolutionary Synthesis" (Wilson, 2010), but this synthesis was formulated just prior to the beginnings of explicitly evolutionary approaches to modern cognition.

In this paper we explore two such transitions. The first, discussed in section 2, is the origin of a richer, post-*Pan*, post-*Australopithecine* culture as early as 2.2 million years ago (Harmand et al., 2015). The second, discussed in section 3, is the explosion of creative culture in the Middle/Upper Paleolithic. Section 4 summarizes simulations carried out using an agent-based model aimed at investigating whether the proposed mechanisms do in fact enhance the capacity for cultural evolution as proposed.

## A First Cognitive Transition

We begin with the archaeological and anthropological evidence for a second transition, followed by our proposed cognitive explanation, and a comparison to other proposals.

### Evidence for a First Cognitive Transition

The minds of *Australopithecus* and earliest *Homo* have been referred to as *episodic* because there is no evidence that their experience deviated substantially from the present or very near-time moment of concrete sensory perceptions. Their archaeological record of simple stone (and some bone and antler) implements indicates that they encoded perceptions of events in memory—an information-handling capacity that supplies "timely information to the organism's decision-making systems" (Klein et al., 2002, p. 306)—but had little voluntary access to memories without external cues, which meant minimal innovation and artifact variation.

This is reflected in the early archaeological record, beginning with stone tools from Lomekwi 3 West Turkana, Kenya, 3.3 mya (Harmand et al., 2015), and characterized by opportunism in highly restricted environments (Braun et al., 2008). Tools were technologically on par with those of modern chimpanzees (Byrne, 2005; Blackwell & d'Errico, 2001; see Read (2008) and Fuentes (2015) for cognitive considerations of chimpanzee toolmaking). These tools also lack evidence of symbolism (d'Errico et al., 2003), and were transported relatively short distances across landscapes (Potts, 2012). While nut-cracking and other simple tool use outside Homo may involve the sequential chaining of actions, and thus the sequential chaining of the mental representations underlying these actions, outside Homo this kind of processing does not occur with sufficient frequency or diversity to cross the threshold to to engage in abstract thought (see Gabora & Steel, 2017 for a mathematical model of what is needed for this threshold to be crossed). Thus, the evidence suggests that the ability of early Homo to voluntarily shape, modify, or practice skills and actions was at best negligible, and they could not invent or refine complex actions, gestures, or vocalizations.

Early *Homo* evolved into a variety of forms, including *H. erectus*, dating between 2.8 - 0.3 million years ago (Villmoare et al. 2015). Multiple lines of evidence suggest a shift away from biology and towards culture as the primary means of adaption in this lineage, attended by significant cultural elaboration. Having expanded out from Africa as early as 2 mya, *Homo* constructed tools involving more production steps and more varied raw materials (Haidle, 2009), imposed symmetry on tool stone (Lepre et al., 2011), used and controlled fire (Goren-Inbar et al., 2004), ranked moderately high among predators (Plummer, 2004), crossed stretches of open water up to 20 km (Gibbons 1998), ranged as far north as latitude 52◦ (Parfitt et al., 2010), revisited campsites possibly for seasons at a time, sometimes built shelters (Mania & Mania 2005), and transported tool stone over greater distances than their predecessors (Moutsou, 2014).

It is widely believed that these signs of a culture richer than that of *Pan* or *Australopithecus* c. 1.7 mya reflect a transition in cognitive and/or social characteristics significantly beyond the small-space, short-time episodic 'bubble' of earlier minds. While the cranial capacity of *Homo erectus* was approximately 1,000 cc—about 25% larger than that of *Homo habilis*, at least twice as large as that of living great apes, and 75% that of modern humans (Aiello, 1996)—brain volume alone cannot explain these developments, which imply an important cognitive transition.

**Background to Proposed Cognitive Mechanism Underlying First Transition**

Because the cognition of *Homo habilis* was primarily restricted to the "here and now" of the present moment, Donald (1991) refers to it as an *episodic mode* of cognitive functioning. He proposed that with the enlarged cranial capacity of later early *Homo*, the hominin mind underwent a transition to a new mode of cognitive functioning made possible by the onset of what he calls a *self-triggered recall and rehearsal loop*, which we abbreviate STR. STR enabled hominins to voluntarily retrieve stored memories independent of environmental cues (sometimes referred to as 'autocuing') and engage in RR (representational redescription) and the refinement of thoughts and ideas. Donald referred to this new kind of mind as the 'mimetic mind' because it could act out or 'mime' events that occurred in the past or that could occur in the future, thereby not only temporarily escaping the present, but through mime or gesture communicating the escape to others.

STR also enabled attention to be directed away from the external world toward ones' internal representations, which paved the way for abstract thought. We use the term *abstract thought* to refer to the processing of previously assimilated experiences, as in occurs counterfactual thinking, planning, or creativity, as opposed to direct perception of the concrete 'here and now' (for a review of abstract thought, see Barsalou, 2005). Note that in much of the cultural evolution literature, social learning is contrasted with individual learning, which involves learning for oneself, and novelty is attributed to things like copying error (e.g., Henrich & Boyd, 2002; Mesoudi, Whiten & Laland, 2006; Rogers, 1988). Abstract thought and creativity, if mentioned at all, are equated with individual learning. However, they are not the same thing. Individual learning deals with obtaining pre-existing information from the environment through non-social means (e.g., learning to predict weather patterns by watching the clouds). In contrast, abstract thought involves mental processing of internally derived contents, and when this results in the generation of useful or pleasing ideas, behavior, or artifacts that did not previously exist, it is said to be creative. Thus, in the case of individual learning, the information comes from the external world, while in the case of abstract thought, it is internally generated. Indeed, there is increasing recognition of the extent to which creative outcomes are contingent upon internally driven incremental/iterative processing (Basadur, 1995; Chan & Schunn, 2015; Feinstein, 2006; Gabora, 2017).

Note that Donald's explanation focuses on neither technical nor social abilities but on a cognitive trait that could facilitate both. STR enabled systematic evaluation and improvement of thoughts and motor acts by adapting them to new situations, resulting in voluntary rehearsal and refinement of skills and artifacts. STR also broadened the scope of social activities to include pantomime and re-enactive play.

**Proposed Cognitive Mechanism Underlying First Transition**

Leaving aside alternatives to Donald's proposal until the end of this section, for now we note that although Donald's explanation seems reasonable so far as it goes, it does not explain why larger brain size enabled STR. What was taking place at the at the level of associative memory that made STR possible? In what follows, we contextualize Donald's (1991) generally well received but sketchy theoretical schema in more current literature. Building on Donald's proposal that the cognitive abilities of modern Homo are due to an accumulation of modes of representation post-Pan starting with the onset of STR, we will ground the concept of STR in a neural level account of the mechanisms underlying cognitive flexibility and creativity (Gabora, 2010; Gabora & Ranjan, 2013).

We start by summarizing a few well-known features of associative memory. Each neuron is sensitive to a primitive stimulus attribute, or microfeature, such as lines of a particular orientation, or sounds of a particular pitch. Items in memory are *distributed* across cell assemblies of such neurons; thus each neuron participates in the encoding of many items. Memory is also *content-addressable:* there is a systematic relationship between the content of an item and the neurons that encode it; thus, items that share microfeatures may be encoded in overlapping distributions of neurons.

We propose that, while in and of itself increased brain volume does not explain the origin of BM, larger brains enabled a transition from more coarse-grained to more fine-grained memory. The smaller the number of neurons a brain has to work with, the fewer attributes of any given item it can encode, and less able it is to forge associations on the basis of shared attributes. Conversely, the evolution of a more fine-grained memory meant that representations could be encoded in more detail, i.e., distributed across larger sets of cell assemblies containing more neurons. Since the memory organization was content addressable that meant more ways in which distributed representations could meaningfully overlap.

Greater overlap enabled more routes by which one memory could evoke another. This in turn made possible the onset of STR, and paved the way for the capacity to engage in recursive recall and streams of abstract thought, and a limited kind of insight (Gabora, 2002, 2010; Gabora & Ranjan, 2013). To take a simple example, the reason that the experience of being accidentally punctured by a thorn could potentially play a role in the invention of an arrowhead is that both the thorn wound and hunting experiences involve overlap in the set of relevant attributes (i.e., "pointed", "flesh", "tear"), and thus overlap of activated cell assemblies.

Representations could now be reprocessed until they achieved a form that was acceptably consistent with existing understandings or sufficiently enabled goals and desires to be achieved (Gabora, 1998). This scenario provides a plausible neural-level account of Donald's (1991) proposal that abstract thought was a natural consequence of possessing a self-triggered recall and rehearsal loop, which was made possible by the increase in brain size at this time.

**Comparison to Other Theories**

We now compare this theory to prominent theories concerning the cognitive underpinnings of the origin of rich, post-*Australopithecine* culture.

Some theories attribute the origins of rich, post-*Australopithecus* culture to social factors. Foley and Gamble (2009) place the emphasis on enhanced family bonding and the capacity for a more focused style of concentration, further enhanced by controlled use of fire by at least 400,000 years ago. Wiessner (2014) suggests that fire not only enabled the preparation of healthier food, but by providing light after dark, facilitated playful and imaginative social bonding. Others emphasize an extrication from biologically based to culturally based kinship networks (Leaf & Read, 2012; Read, 2012; Read & van der Leeuw, 2015). We believe that these social explanations are essentially correct, but that they have their origin in cognitive changes, which altered not only social interactions but interactions with other facets of human experience as well.

Our proposal bears some resemblance to Hauser et al.'s (2002) suggestion that what distinguishes human cognition from that of other species is the capacity for recursion, Penn et al.'s (2008) concept of relational reinterpretation, and Read's (2009) claim that relational concepts and recursive reasoning allowed for a conceptually based system of social relations but may have evolved in conjunction with non-social activities such as toolmaking. While our proposal is consistent with this, it goes further, by grounding the onset of recursive reasoning in a transition in the structure of associating memory. Read suggests that recursive reasoning was made possible by larger working memory, while we argue that larger working memory in and of itself is not useful; it must goes hand-in-hand with (and indeed is a natural byproduct of) more fine-grained memory. As a simple example, let us suppose that a hominid with a coarse-grained memory increased its working memory from being able to think only of one thing at a time (e.g., a thorn) to two (e.g., a thorn and the sun). This would generally be a source of confusion. However, if it held only one thing in mind at a time but encoded it in richer detail (e.g., incorporating attributes of a thorn such as 'sharp', 'pointy', 'thin', and so forth), it could forge meaningful associations with other items based on these attributes (e.g., other sharp things or pointy things).

Our proposal also bears some resemblance to Chomsky's (2012) concept of 'merge'. However, while 'merge' is described as the forging of associations between items that are extremely similar, or that co-occur in time or space, STR can additionally forge associations between items that are related by as few as a single attribute, and do so recursively such that the output of one such operation is the input for the next, and reliably, such that encodings are modified in light of each other in the course of streams of thought (Gabora, 2002, 2013, 2017, 2018). (Detailed examples—including the invention of a fence made of skis on the basis of the attributes 'tall', 'skinny' and 'sturdy' (Gabora, 2010), and the generation of the idea of a beanbag chair on the basis of the single attribute 'conforms to shape' (Gabora, 2018)—are provided elsewhere.)

Thus, while merge forges associations based on overall similarity, for STR the memory must be sufficiently fine-grained (i.e., items must be encoded in enough detail) that the associative process can operate on the basis of *specific attributes* to which specific neurons are tuned. Thus our proposal (but not 'merge') offers a causal link between brain size and cognitive ability, i.e., more neurons means they can be tuned to a wider range of attributes and thereby form more associations on the basis of shared attributes.

Mithen's (1996) model features the accumulation and overlap of a variety of intelligence modules. Although in its details his model runs rather counter to much current thinking including our own, his focus on cognitive fluidity and creativity influenced the model proposed here.

## A Second Cognitive Transition

As with the first transition, we begin with the archaeological and anthropological evidence for a second transition, followed by our proposed cognitive explanation, and finally a comparison to other proposals.

### Evidence for a Second Transition

The African archaeological record indicates that another significant cultural transition occurred approximately 100,000 years ago, bearing many of the material correlates of BM. Though defining BM is somewhat difficult (d'Errico et al., 2005; Shea 2011), prehistorians generally agree that BM is evidenced in the archaeological record by a spatially and temporally quite disparate suite of artifacts and characteristics including (a) artifacts indicating personal symbolic ornamentation (d'Errico et al., 2009), (b) elaborate burial sites indicating ritual (Hovers, Ilani, Bar-Yosef, & Vandermeersch, 2003) and possibly religion (Rappaport, 1999), (c) a radical proliferation of tool types that better fit tools to specific tasks (McBrearty & Brooks, 2000), (d) 'cave art', i.e., representational imagery featuring depictions of animals (Pike et al., 2012) and human beings (Nelson, 2008), (e) complex hearths and highly structured use of living spaces (Otte, 2012), (f) extensive use of bone and antler tools, sometimes with engraved designs, and (g) calorie-gathering intensification that included widespread use of aquatic resources (Erlandson, 2001). BM spread out of Africa sometime after 100,000 years ago, and was present in Sub-Himalayan Asia and Australasia over 50,000 years ago (Mulvaney & Kamminga, 1999) and Continental Europe not long thereafter (Mellars, 2006).

Whether this archaeological record reflects a genuine transition resulting in BM is hotly debated because claims to this effect are based on the European Paleolithic record, and largely exclude the lesser-known African record (Fisher & Ridley, 2013). Many artifacts associated with a rapid transition to BM 40,000-50,000 years ago in Europe are found in the African Middle Stone Age tens of thousands of years earlier, which pushes the cultural transition more closely into chronological alignment with the transition to anatomical modernity between 200,000 and 100,000 BP. Nevertheless, it is clear that modern behavior appeared in Africa between 100,000 to 50,000 years ago, and spread, resulting in displacement of the Neanderthals in Europe (Fisher & Ridley, 2013). Subsequently, the cultures of *Homo sapiens* were radically more open-ended and accumulative, meaning that they could archive effectively infinite amounts of information to be used in adaptation, one of the adaptive advantages of complex culture. Despite a lack of overall increase in cranial capacity, the prefrontal cortex, and more particularly the orbitofrontal region, increased significantly in size (Dunbar 1993), in what was most likely a time of major neural reorganization (Morgan 2013).

**Proposed Cognitive Mechanism Underlying Second Transition**

Given that behaviorally modern humans were demonstrably more creative than any prior hominin (Mithen, 1998), what role could changes at the cognitive level have played in their evolution?

We propose that the cultural explosion of the Middle/Upper Paleolithic came about due to fine-tuning of the biochemical mechanisms underlying the capacity to spontaneously shift between different modes of thought depending on the situation by varying the specificity of the activated memory region. The ability to shift between different modes is referred to as *contextual focus* (CF) because it requires the capacity to focus or defocus attention in response to contextual factors (Gabora, 2003), such as the audience, or level of danger, or goals, which may shift minute by minute if goals are broken into subgoals. Focused attention is conducive to *analytical thought* (Agnoli, Franchin, Rubaltelli, & Corazza, 2015; Vartanian, 2009; Zabelina, 2018). In analytic thought, the activation of memory is constrained enough to hone in and mentally operate on only the relevant aspects of the contents of thought. In contrast, by diffusely activating a wide region of memory, defocused attention is conducive to *associative thought*; it enables more obscure (but potentially relevant) aspects of the situation to come into play. This greatly enhances the potential for insight, i.e., the forging of obscure but useful or relevant connections.

Once the products of one mode of thought could become 'ingredients' for the other, they could reflect on the contents of their mind not just from different perspectives but at different levels of granularity, from basic level concepts (e.g., deer) up to abstract concepts (e.g., animal) and down to more detailed levels (e.g., legs), as well as conceive of their interrelationships. All this was necessary in order to have a need to come up with names for these things, i.e., develop complex languages. Thus, it is proposed that CF paved the way for not just language but a range of cognitive abilities considered by anthropologists to be diagnostic of BM. Note that associative thought is useful for breaking out of a rut, but would be risky without the ability to reign it back in; basic survival related tasks may be impeded if everything is reminding you of everything else. Therefore, it seems reasonable that it would take considerable time to fine-tune the mechanisms underlying the capacity to spontaneously shift between these two processing modes such that one retained the benefits of escaping local minima without running the risk of being perpetually side-tracked. The time needed to fine-tune this could potentially be the explanation for the lag between anatomical and BM.

**Comparison to Other Theories**

We now review some prevailing hypotheses for how and why BM and its underlying intellectual capacities arose.

Our proposal is superficially similar to the idea that what distinguishes human cognition from that of other species is our capacity for dual processing (Evans, 2008; Nosek, 2007). Dual processing posits that humans engage in not just a primitive implicit Type 1 mode for free association and fast “gut responses", but also an explicit Type 2 mode for deliberate analysis. However, while dual processing makes the split between older, more automatic processes and newer, more deliberate processes, CF makes the split between an older associative mode based on relationships of correlation and a newer analytic mode based on relationships of causation. We propose that although earlier hominids relied on the older association-based system, because their memories were coarser-grained, there were fewer routes for meaningful associations, so there was less associative processing of previous experiences. Therefore, items encoded in memory tended to remain in the same form as when they were originally assimilated; rather than engaging in associative or analytic processing of previously assimilated material, there was greater tendency to focus on the here and now.

Thus, while dual processing theory attributes abstract, hypothetical thinking to the more recent Type 2 mode, according to the CF hypothesis it is possible in either mode but differs in character in the two modes (logically constructed arguments in the analytic mode versus flights of fancy in the associative mode). The CF hypothesis is rooted in a distinction in the creativity literature between (1) associative (divergent) processes said to predominate during idea generation, and (2) analytic convergent processes said to predominate during the refinement, implementation, and testing of an idea (Finke, Ward, & Smith, 1992). (See Sowden, Pringle, & Gabora (2014), for a comparison and discussion of the relationship dual processing theory and dual theories of creativity; see Gabora, 2018 for discussion of the distinction between associative versus divergent thought).

To see how the onset of CF could give rise to open-ended cultural complexity, recall the previously-mentioned properties of associative memory: distributed representation, coarse coding, and content addressability. Each thought may activate more or fewer cell assemblies depending on the nature of the task at hand. Focused attention is conducive to analytic thought because memory activation is constrained enough to zero in and operate on the most defining properties. Defocused attention, by diffusely activating a diversity of memory locations, is conducive to associative thought; obscure (but potentially relevant) properties of the situation come into play (Gabora, 2000, 2010). Thus, while in an analytic mode of thought the concept TOOL might only activate ‘hand axe’, in an associative mode of thought, all sorts of items in ones’ environment might potentially be used as a tool depending on what one wants to accomplish. Once it was possible to shift between these modes of thought, cognitive processes requiring either analytic thought, associative thought or both could be carried out more effectively, and the fruits of one mode of thought could become ingredients for the other mode, thereby facilitating the forging of a richly integrated creative internal network of understandings about the world and one's place in it, which we refer to as a *worldview*[1]. This in turn set in motion behavioral modernity. Thus, the notion that diffuse activation is conducive to associative thought while activation of a narrow receptive field is conducive to analytic thought is consistent with the architecture of associative memory, and suggests a means by which CF made possible the capacity to stay on task, yet, when needed, forge unusual yet relevant connections. Language enhanced not just the ability to communicate and collaborate (thereby accelerating the pace of cultural innovation), but also the ability to think ideas through for oneself and manipulate them in a controlled, deliberate manner.

In short, we propose that the emergence of a self-organizing worldview required two transitions, as illustrated schematically in Figure 1. The onset of STR over 2 mya (as discussed above) allowed rehearsal and refinement of skills and made possible minor modifications of representations. The onset of CF approximately 100,000 years ago made it possible to forge larger bridges through conceptual space that paved the way for innovations specifically tailored to selective pressures. It enabled a cultural version of what Gould and Vrba (1982) termed *exaptation*, the phenomenon wherein an existing trait is co-opted for a new function (Gabora, Scott, & Kauffman, 2013). Exaptation of representations and ideas dramatically enhanced the ability to, not just cope with the technological and social spheres of life, but develop individualized perspectives and unique worldviews conducive to fulfilling complementary social roles. This increase in cognitive variation provided the raw material for better adaptive fit to selective pressures.

Mithen (1996) proposed that the integration of previously-compartmentalized intelligence modules–specialized for natural history, technology, socialization and language–lay at the heart of BM. That integration is said to have enabled *cognitive fluidity*: the capacity to combine concepts and adapt ideas to new contexts, and thereby explore, map, and transform conceptual spaces across different knowledge systems. Fauconnier and Turner (2002) emphasize the benefit of cognitive fluidity for the capacity to draw and understand analogies. Our proposal is consistent with these explanations, but goes beyond them by showing how conceptual fluidity arises naturally as a function of the capacity to, when needed, shift to a more associative mode of processing.

There are many variants of the theory BM reflects onset of sophisticated language abilities. Corballis (2011) suggests that this may have entailed a transition from a predominantly gestural to a vocal form of communication. Although the ambiguity of the archaeological evidence makes it difficult to know exactly when language began (Davidson & Noble 1989; Christiansen & Kirby, 2003; Hauser, Chomsky, & Fitch, 2002),'it is widely believed—based on stone tool symmetry and complexity of manufacture—that as long ago as c. 1.7 million years Homo used gestural and prelinguistic vocalization communications that would have shared some organizational similarities with modern humans insofar as they differed significantly from other primate communications". The evolution of grammatically- and syntactically-modern language is generally placed (depending on whether one is observing it in Africa, sub-Himalayan Asia or Western Eurasia) after about 100,000 years ago, around the start of the Upper Palaeolithic (Bickerton, 2014; Dunbar, 1993; Tomasello, 1999).

---

[1] Our concept of worldview is closely aligned with what Read (2013) refers to as a 'cultural idea system'.

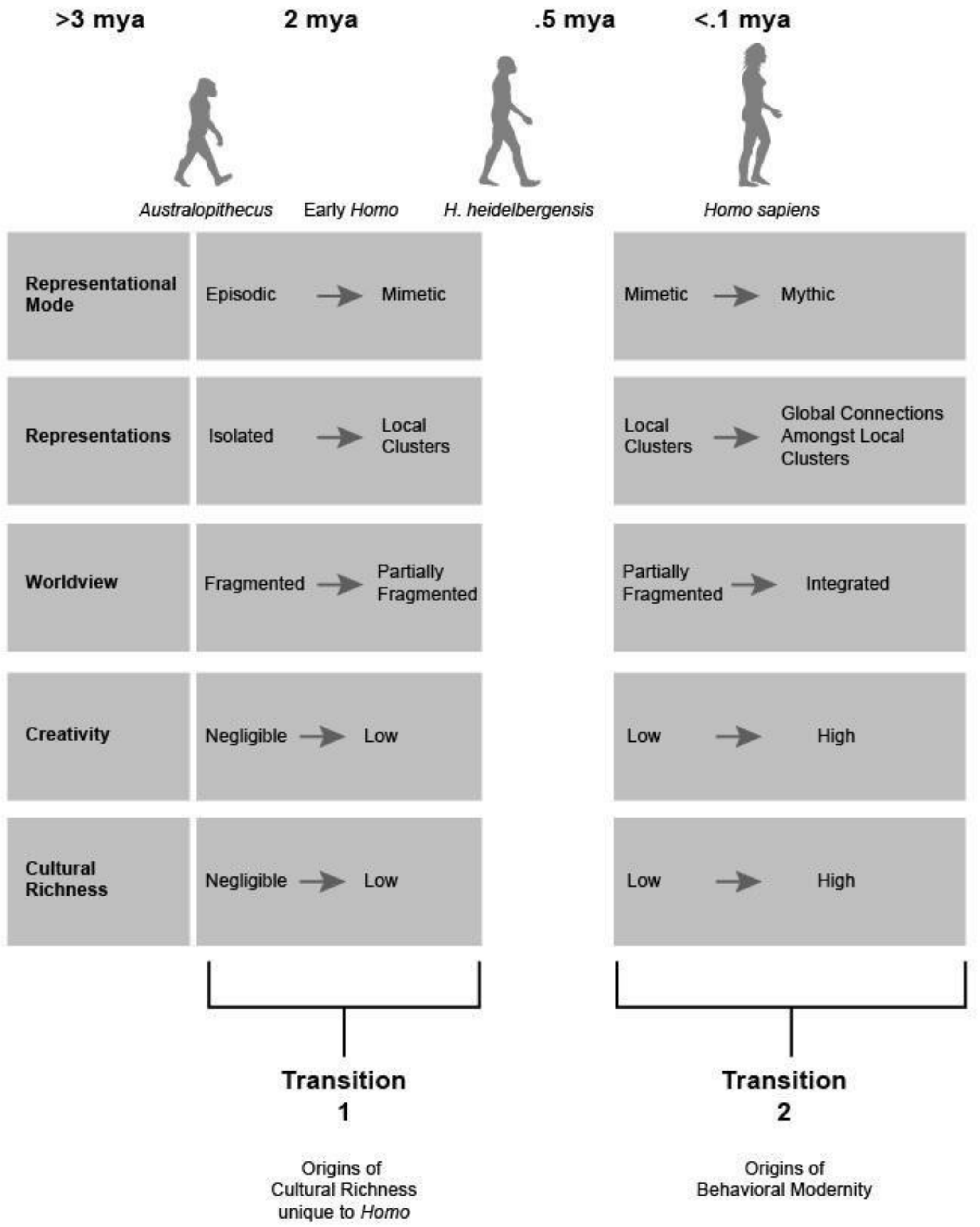

*Figure 1. Schematic illustration of the proposed cognitive transitions resulting in behavioral modernity. Over time cognitive features undergo transitions from earlier to later states; different features are involved in transitions 1 and 2, which are separated by over a million years, with cognition evolving in a mosaic fashion.*

Bickerton (2014) proposes that BM entailed a series of stages (sensu Szathmary, 2015), though he focuses more specifically on language. In his view, open-ended cultural evolution began with selection for brain mechanisms underlying cognitive reorganization, ‘offline thinking’, and the elaboration and ‘ratcheting’ (cf Tomasello, Kruger & Ratner, 1993) of cultural information. Deacon (1997) emphasizes onset of the capacity to internally represent complex, abstract, internally coherent systems of meaning using symbols. Carstairs-McCarthy (1999) suggests that some form of syntax was present in the earliest languages, but most of the later elaboration, including recursive embedding of syntactic structure, evolved with BM. It is widely accepted that syntax constituted a crucial step toward BM, as it made it possible to state more precisely how elements are related, and embed them in other elements (Bickerton & Szathmáry, 2009). Thus, syntax enabled language to become general-purpose and applied in a variety of situations, highly unlike the situation-specific communication that has been observed in other species such as vervet monkeys. Donald (1991) proposed a transition in the mode of representation, enabling the capacity for narrative myth, as the underwrite of BM. Once again, our proposal is consistent with the idea that complex language abilities lie at the heart of BM, but because STR followed by CF would have enabled hominids to not just recursively refine and modify thoughts but consider them from different perspectives at different hierarchical levels, it would have stage the stage for complex language and facilitated the weaving of experiences into stories, parables, and broader conceptual frameworks, thereby integrating knowledge and experience (see also, Gabora & Aerts, 2009).

Another proposal is that recursion, featuring "the [cognitive] creation of sequences or [thought] structures of unbound length or complexity" enabled mental time travel, distinctly-human cognition, and BM (Corballis, 2011, p. 5-6; see also Suddendorf et al., 2009). Proponents note the limited use of recursion in *Pan*, for instance, but its centrality in modern human cognition. Corballis suggests that recursion allowed for self-actuated recall of past episodes (analogous to Donald's 'autocuing') and cognitive models of possible futures, resulting in not just deeper individual bonding and information sharing but also "deeper levels of Machiavellian intrigue" (Corballis, 2011, p. 222). For reasons outlined earlier, we believe that recursive reasoning came about well before BM, though the ability to shift between different modes of thought using CF would have brought on the capacity to make vastly better use of it.

Another proposal is that BM arose due to onset of the capacity to model the contents of other minds, sometimes referred to as 'Theory of Mind' (Tomasello, 2014). Tomasello further proposes that this resulted in "shared intentionality", involving exchange of knowledge and goals, and potentially accentuation of group concerns over those of the individual, paving the way for social selection favoring cohesive groups.

This explanation for BM is somewhat underwritten by recursion—in other words, the mechanism that allows for recursion is required for modeling the contents of other minds—but the emphasis is on the social impact of recursion, rather than the capacity for recursion itself. Other social-ecological theories emphasize different factors. Whiten (2011) emphasizes a gradual increase in the complexity of social learning processes leading to the generation and ratcheting of richer, more diverse cultural traditions and extrasomatic culture. Foley and Gamble (2009) examined the 'ecology of [hominin] social transitions'; it is in their fifth transition (after 200,000 BP) that BM 'appears'. Our proposal is consistent with explanations that stress the onset of social abilities, but places these explanations in a broader framework by suggesting a mechanism that aided not just social skills but other skills (e.g., technological) as well.

While most of these explanations are correct insofar as they go, we suggest that none of them go sufficiently to the root of the matter. As Carl Woese wrote of science at large "...sometimes [there is] no single best representation... only deeper understanding, more revealing and enveloping representations," (Woese 2004:173). We propose that the second cognitive transition necessary for cumulative, adaptive, open-ended cultural evolution was the onset of CF, because once hominids could adapt their mode of thought to the situation they were in, and sculpt the output of such thought processes by subjecting them to different perspectives, and different levels of analysis, their initially fragmented mental models of their world could be woven into more coherent mental models of their world—i.e., worldviews—which facilitated not just conceptual fluidity, creative problem-solving, and survival, skills but also interactive social exchange and more complex social structures. We add that the explanation proposed here is the only one we are aware of that grew out of a synthesis of archaeological and anthropological data with theories and research from both psychology and neuroscience. In addition, it is supported by computational simulations, to which we turn next.

## 5. Simulation of Two Cultural Transitions

We have reviewed the evidence for two hypotheses: (1) the earliest signs of culture were due to the onset of STR, which enabled representational redescription and abstract thought, and (2) the cultural explosion of the Middle-Upper Paleolithic was due to the onset of CF. We now summarize support for the hypothesis that these abilities played vital roles in the arrival of behavioural modernity and cultural evolution obtained using an agent-based model of cultural evolution referred to as "EVOlution of Culture", abbreviated EVOC. EVOC uses neural network based agents that (1) invent new ideas, (2) imitate actions implemented by neighbors, (3) evaluate ideas, and (4) implement successful ideas as actions. EVOC is an elaboration of Meme and Variations, or MAV (Gabora, 1995), the earliest computer program to our knowledge to model not just cultural transmission but cumulative, adaptive, cultural evolution.[2]

The goal behind EVOC (and MAV) was to distill the underlying logic of cultural evolution. As such, it is a vastly simplified model, much simpler than models of language evolution (e.g., Steels, 2012). Agents do not evolve in a biological sense—they neither die nor have offspring—but do in a cultural sense, by cumulatively modifying each others' ideas for actions. Results obtained with this model may or may not tell us something about what is going on in the real world, but it allows us to vary one parameter while holding others constant and thereby test otherwise untestable hypotheses. This approach is particularly useful for studies at the interface between anthropology and psychology due to the sparseness of the pre-modern archaeological record. Although methods for analyzing these remains are becoming increasingly sophisticated, they cannot always distinguish amongst competing theories. Thus, computational models can be particularly valuable, providing a means of assessing the feasibility of theories concerning the origins of behaviorally modern cognition.

### 4.1 The Computational Model

We summarize the architecture of EVOC in sufficient detail to explain our results; for details see (*e.g.,* Leijnen & Gabora, 2009).

**Agents.** Agents consist of (1) an auto associative neural network, which encodes ideas for actions and detects trends in what constitutes a fit action, (2) a 'perceptual system', which carries out the evaluation and imitation of neighbors' actions, and (3) a body, consisting of six body parts which implement actions. The neural network is composed of six input nodes and six corresponding output nodes that represent concepts of body parts (LEFT ARM, RIGHT ARM, LEFT LEG, RIGHT LEG, HEAD, and HIPS), as well as hidden nodes that represent more abstract concepts (LEFT, RIGHT, ARM, LEG, SYMMETRY, OPPOSITE, and MOVEMENT). Input nodes and output nodes are connected to hidden nodes of which they are instances (*e.g.,* LEFT ARM is connected to LEFT.) Activation of any input node activates the MOVEMENT node. Same-direction activation of symmetrical input nodes (*e.g.,* upward motion–of both arms) activates the SYMMETRY node. Further details concerning the neural network are provided in appendix A.

[2] The approach can thus be contrasted with computer models of how individual learning affects biological evolution (e.g., Higgs 2000; Hinton & Nowlan 1987; Hutchins & Hazelhurst 1991).

**Invention.** An idea for a new action is a pattern consisting of six elements that dictate the placement of the six body parts. Agents generate new actions by modifying their initial action or an action that has been invented previously or acquired through imitation. During invention, the pattern of activation on the output nodes is fed back to the input nodes, and invention is biased according to the activations of the SYMMETRY and MOVEMENT hidden nodes. (Were this not the case there would be no benefit to using a neural network.) To invent a new idea, for each node of the idea currently represented on the input layer of the neural network, the agent makes a probabilistic decision as to whether the position of that body part will change, and if it does, the direction of change is stochastically biased according to the learning rate. If the new idea has a higher fitness than the currently implemented idea, the agent learns and implements the action specified by that idea.

**Imitation.** The process of finding a neighbor to imitate works through a form of lazy (non-greedy) search. The imitating agent randomly scans its neighbors, and adopts the first action that is fitter than the action it is currently implementing. If it does not find a neighbor that is executing a fitter action than its own current action, it continues to execute the current action.

**Evaluation: The Fitness Function.** Fitness was evaluated using an adaptation of the Royal Roads fitness function (Forrest & Mitchell, 1993). Definitions of terms used in the evaluation of the fitness of an action are provided in Table 1. The first fitness function is determined by 45 templates. The second fitness function is constructed analogously but with different sub-actions. The templates can be thought of as defining the cultural significance of types of sub-actions (such as dance steps).

Each template $T^i$ consists of six components, one for each body part (*i.e.,* $T^i = t^i_j$; $j = 1..6$). Each body part can be in a neutral position (0) , up (1), down (-1), or an unspecified position (*). Six examples of templates are provided in Table 2. For example, in template $T^i = *, 1, -1, *, *, 0$, the left arm is up (LA:1), the right arm is down (RA:-1), the hips are in the neutral position (HP:0), and the positions of other body parts is unspecified (HD:*, LL:*, and RL:*). The templates provide constraints, as well as flexibility with respect to what constitutes a fit action. For example, in an optimally fit action, the head must be in the neutral position (in $T^i$ the first component is 0) but the positions of other body parts can vary).

Details of the calculation of the fitness of an action are provided in appendix B. The fitness functions are difficult to solve because they are rugged, consisting of many peaks and valleys; hill-climbing is not guaranteed to lead to an optimal solution. There are multiple fitness peaks, that must be crossed before reaching the plateau. For example, consider the fitness function given in Table 2. The action 0,0,0,0,0,0 has a fitness of 6. An agent may move on from this action to find an actions that fits the third order templates with a fitness of 31, *e.g.,* $F(D)$ : {1, 1, 1, 1, 1, 0} = 3 + 3 + 3 + 3 + 3 + 3 + 3 + 3 + 3 + 3 + 1 = 31. Midway through a run (at iteration 50) the fitness function changes to test the ability to adapt to a sudden change in task constraints or environment.

*Table 1. Definitions and examples of terms.*

| Term | Definition | Example |
|---|---|---|
| Body Part | Component of agent other than neural network. | Left Arm (LA) |
| Sub-action | Set of six components that indicates position of 6 body parts. Each can be in a neutral (0), up (1), or down (-1) position. | {HD:0, LA:1, RA:-1, LL:1, RL:0, HP:-1; This sub-action is abbreviated 01-110-1} |
| Action | One or more sequential sub-actions. | {{01001-1}, {-10-1-111}} |
| Template | Abstract or prototypical format for a sub-action. Position of a body part can be unspecified (*). | {HD:0, LA:*, RA:1, LL:*, RL:1, HP:-1} |

*Table 2. A partial set of the templates used in the first fitness function.*

| | |
|---|---|
| $T^1 = \{0, *, *, *, *, *\}$ | $T^{24} = \{1, *, *, 1, 1, *\}$ |
| $T^2 = \{*, 0, *, *, *, *\}$ | $T^{25} = \{1, *, 1, *, 1, *\}$ |
| $T^3 = \{*, *, 0, *, *, *\}$ | $T^{26} = \{1, *, 1, 1, *, *\}$ |

**Learning.** Invention makes use of the ability to learn trends and respond adaptively to them. Knowledge acquired through the evaluation of actions is translated into educated guesses about how to invent fit actions. For example, an agent may learn that symmetrical movement tends to be either beneficial or detrimental, and bias the generation of new actions accordingly.

**A Typical Run.** Fitness and diversity of actions are initially low because all agents are initially immobile, implementing the same action, with all body parts in the neutral position. Soon some agent invents an action that has a higher fitness than immobility, and this action gets imitated, so fitness increases. Fitness increases further as other ideas get invented, assessed, implemented as actions, and spread through imitation. The diversity of actions increases due to the proliferation of new ideas, and then decreases as agents hone in on the fittest actions. Thus, over successive rounds of invention and imitation, the agents' actions improve. EVOC thereby models how "descent with modification" occurs in a purely cultural context.

### 5.1 Method

**Modeling Chaining (First Transition).** EVOC has been used to simulate a simple form of STR: the capacity to join representations together sequentially, which we refer to as *chaining* (so as not to convey the impression that it is a full-fledged model of the many ways in which STR could occur).

Since our immediate goal was to investigate the impact of chaining (as opposed to faithfully rendering its underlying mechanisms in humans), in these simulations the capacity for chaining was simply turned on or off as opposed to coming about through the evolution of finer grained memory. Chaining gives agents the opportunity to execute multi-step actions. The agent can keep adding a new sub-action to its current action so long as the most recently-added sub-action is both novel and successful. A sub-action D is considered novel if at least one of its components is different from that of the previous sub-action. It is considered successful if there exists a template $T^i$ such that $\Phi(T^i, D)$ is one, as per equation 1.

$$successful(D) = \begin{cases} true & if\ \exists\ T^i : \Phi(T^i, D) = 1 \\ false & otherwise \end{cases} \tag{1}$$

The fitness of an action consisting of more than one sub-action is obtained by adding the number of sub-actions to the fitness of the last sub-action in the sequence. For example, if the last sub-action of an action is $D$ = [0, 1, −1, 1, −1, 1] and the number of sub-actions is seven, the fitness of the action is $F(D) + 7 = 14 + 7 = 21$. Thus where $c$ is 'with chaining', $w$ is 'without chaining', $n$ is the number of chained sub-actions, the fitness of a chained action, $F_c$, is calculated as per equation 2.

$$F_c = F_w + n \tag{2}$$

An agent can execute an arbitrarily long action so long as it continues to invent successful new sub-actions. In general, the more sub-actions the fitter the action. Chaining is admittedly a simple form of RR, but the goal here was simply to test hypotheses about how the capacity for this kind of (by some definitions) recursive process operating at the individual level affects the dynamics at the societal level.

**Modeling Contextual Focus (Second Transition).** Mathematical models both chaining of CF, and their impact on the global structure of the conceptual idea network or worldview, have been developed (Gabora & Aerts, 2009; Gabora & Steel, 2017), and the model of CF was consistent with experimental data from a study in which participants were asked to rate the typicality of exemplars of a concept for different contexts (Veloz, Gabora, Eyjolfson, & Aerts, 2011). CF was also incorporated into a portrait painting computer program generated artworks that humans preferred over those generated without CF (DiPaola & Gabora, 2009). However, the portrait painting program did not allow investigation of the effect of CF on the evolution of ideas through cultural interaction. Therefore, CF was also modeled using EVOC. In the convergent mode, the current action is only slightly modified to create a new action. In the divergent mode, the current action is substantially modified to create a new action. An agent switches between these modes by modifying its *rate of creative change* (RCC). If the fitness of its current action is low relative to previous actions, RCC increases, causing the agent to shift to a more divergent processing mode conducive to large leaps through the space of possibilities. If action fitness is high relative to that of previous actions, RCC decreases, and the agent shifts to a more convergent mode conducive to minor adjustments. With CF turned off, RCC stays constant throughout the run at 1/6 (*i.e.*, a new action involves change to one of the six body parts). The equation to modify RCC is shown in Equation 3.

$$\Delta RCC = -a(F_{new} - F_{old}) \tag{3}$$

Since at the start of a run previous fitness is undefined, RCC in this case is a function of the current fitness as per Equation 4, where $0 < b < 1$.

$$RCC_{initial} = b^{F_{current}} \tag{4}$$

In the results shown here, $a$ and $b$ were initialized to -0.005 and 0.8 respectively. The implementation of neither chaining nor CF, chaining alone, CF alone, and both chaining and CF simultaneously, are schematically illustrated in panels a, b, c, and d, respectively of Figure 2.

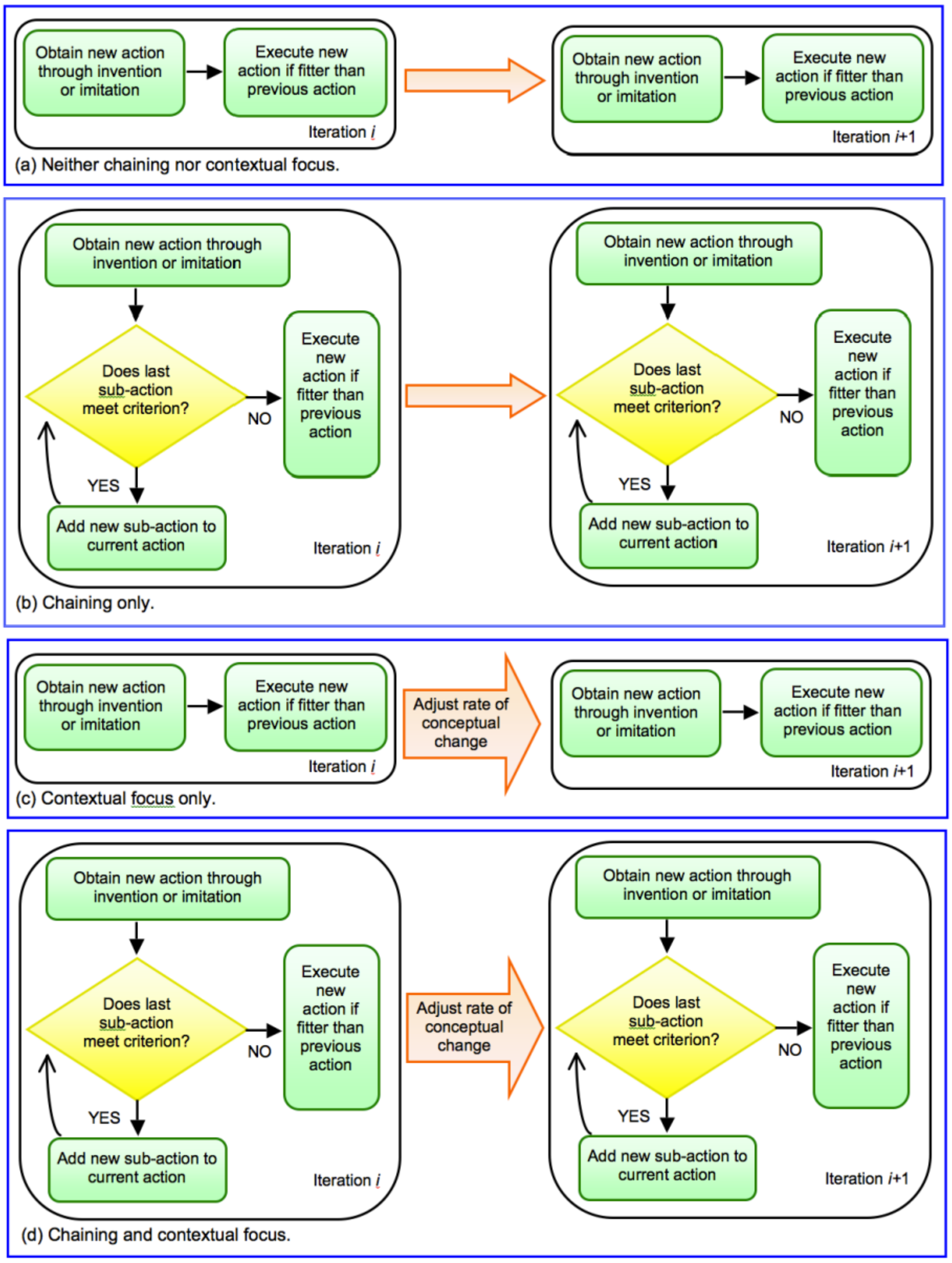

*Figure 2.* *Schematic illustration of the algorithm without chaining or CF, with chaining only, and with both.*

### 5.2 Results and Discussion

The results of incorporating chaining and CF into the method by which agents generated cultural novelty are summarized in Table 3. The results of introducing chaining and CF on the mean fitness and diversity (total number of different actions) of actions across all agents in the society are shown in Figures 3 and 4 respectively.

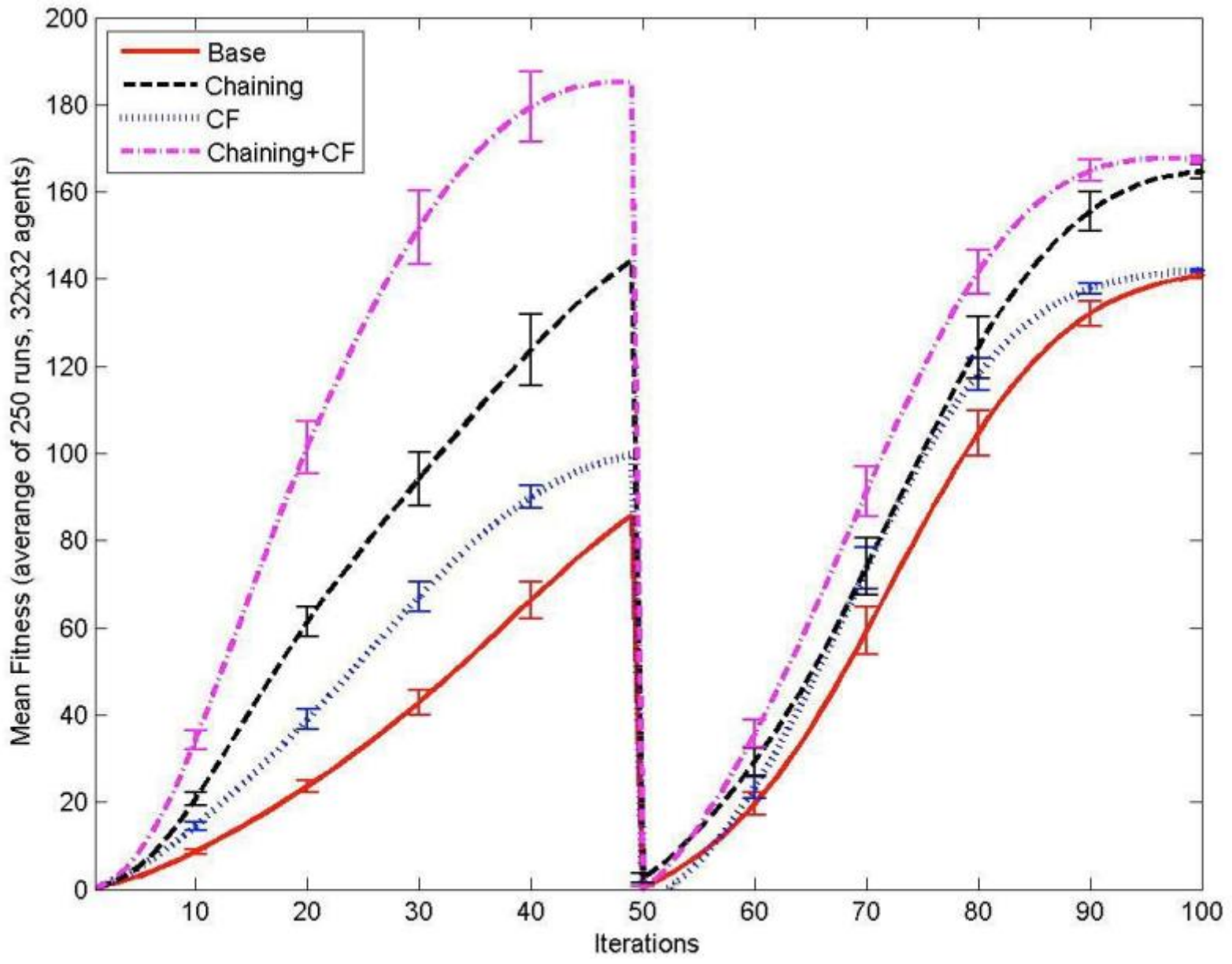

*Figure 3. Mean fitness of cultural outputs across the society with both chaining and CF (red line), chaining only (dashed blue line), and neither chaining nor CF (dotted green line). Data are means of 500 runs. To test the ability to respond to change in the task or environment, there was a change of fitness function at iteration 50. While chaining and CF were both beneficial, the capacity for major changes using CF was ultimately of little value without the ability to make minor refinements using chaining. The fact that CF was only beneficial following exposure to a new fitness function is consistent with its hypothesized role in facilitating new ways of thinking.*

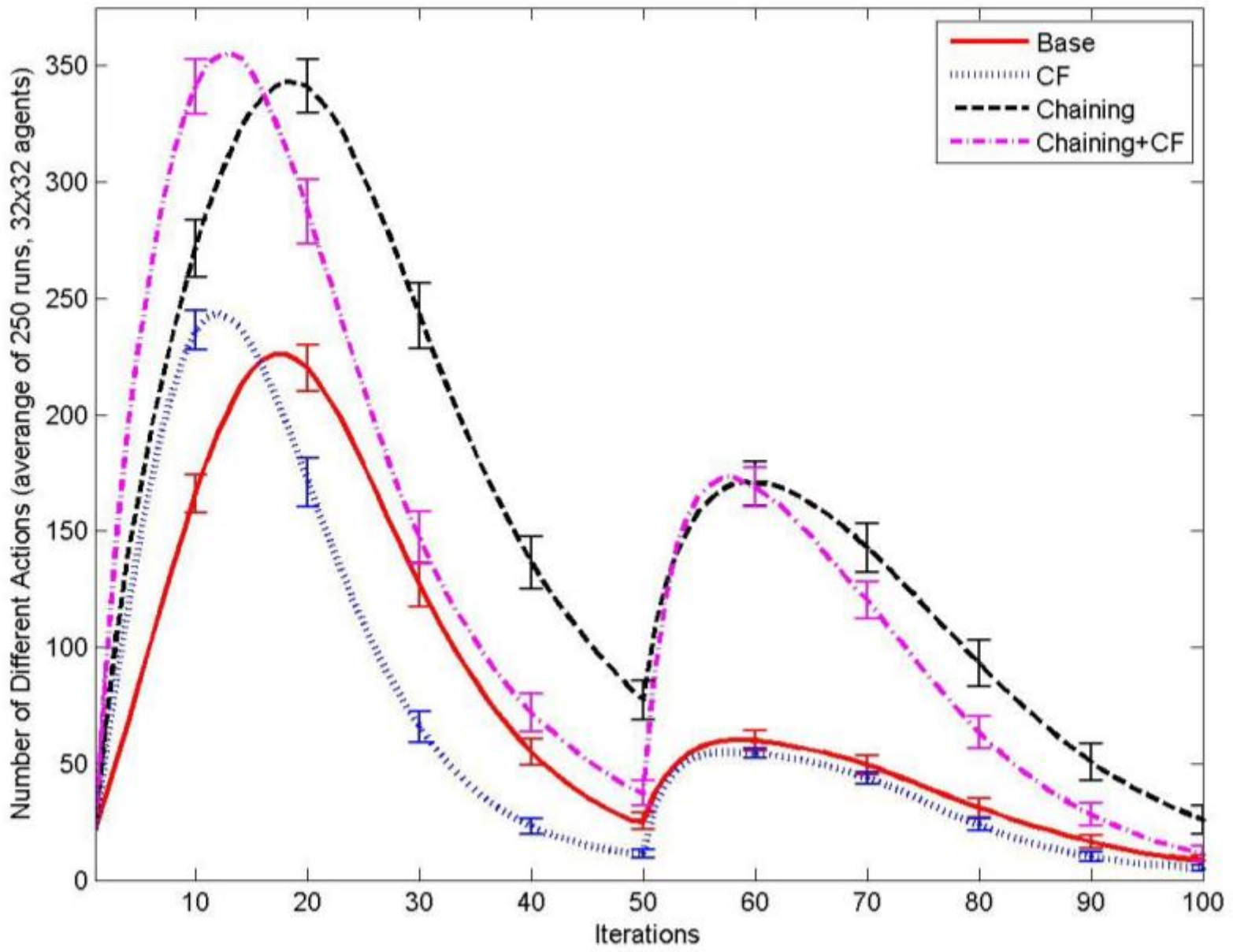

*Figure 4. Diversity of cultural outputs across the society with both chaining and CF (red line), chaining only (dashed blue line), and neither chaining nor CF (dotted green line). Data are means of*

*500 runs. To test ability to respond to change in task or environment there was a change of fitness function at iteration 50. As with fitness, while chaining and CF both increased diversity, their impact was different. While chaining consistently increased diversity, CF tended to exaggerate both the initial increase in diversity early on as the space of possibilities was being explored, and the latter decrease in diversity as agents converged on the fittest outputs. This is consistent with its hypothesized role in adjusting the mode of thought over the course of creative problem solving.*

Chaining and CF both significantly increased the mean fitness of actions. Without chaining, mean fitness quickly reached a plateau; with chaining, the space of possibilities became open-ended, and thus the fitness of cultural outputs could increase indefinitely. This is consistent with the hypothesis that chaining enabled the ratcheting of outputs necessary for cultural change to become an evolutionary process. Inspection revealed that although there is always convergence on optimal actions, without chained actions this set was static because the space of possibilities was finite, thus mean fitness plateaued. On the other hand, with chained actions the space of possibilities was not finite, and the set of optimal actions changed slowly but continuously as increasingly fit actions continued to be found.

CF made a contribution to fitness above that of chaining alone. While chaining increased mean fitness throughout the run, CF was most effective following initial exposure to a new fitness function, i.e., at the beginning of the run or at iteration 50 when the fitness function changed. This supports CF's hypothesized role in responding to radical change. When agents were first exposed to a fitness function, CF increased both the rate at which new possibilities were generated and the rate of convergence on the fittest of these.

Both chaining and CF also significantly increased the diversity, or number of different actions, as shown in Figure 4. Chaining exaggerated both the initial increase in diversity as the space of possibilities was explored, and the subsequent decrease in diversity as agents converged on the fittest actions. As with fitness, CF alone exerted no noticeable effect on diversity once agents had fit actions. However, if chaining was already in place, CF made the inverted-U shaped pattern even more pronounced. The fact that CF had a negligible effect on fitness and diversity of actions unless chaining was already in place is consistent with the hypothesis that chaining arose first and CF arose second.

*Table 3. Summary of agent-based model results.*

| DV | No Chaining, No CF | Chaining Only | CF Only | Chaining + CF |
|---|---|---|---|---|
| Fitness of Actions | Reached plateau | Increased indefinitely (open-ended) | Reached plateau | Increased indefinitely (open-ended) |
| Diversity of Actions | Increased then converged on fittest | Explosive increase, faster convergence | Increased then converged on fittest | More explosive increase, even faster convergence |
| Set of Optimal Actions | Static | Constantly fluctuates as fitter actions found | Static | Constantly fluctuates as fitter actions found |

## General Discussion

We have outlined a speculative but coherent, multilevel explanation for how the the uniquely human capacity for collectively generated, open-ended, adaptive cultural evolution could have come about. Although change occurred in a mosaic fashion in the Homo lineage over a period spanning more than two million years, the resulting overall pattern may be discerned as comprising two significant evolutionary transitions. First, the larger brain of *H. erectus* resulted in finer grained memory with detailed representations, paving the way for rehearsal of actions, refinement of skills, novel associations between closely related items in memory. This enabled STR, escape from episodic proximity, representational redescription (RR), minor improvements in cultural outputs, and a "cultural ratcheting" that expanded the capacity for open-ended cultural evolution. Much later, around 100,000 BP, newly-evolved basal ganglia circuits enabled onset of contextual focus: the ability to shift between convergent and divergent modes of thought, enabling hominins to process information from different perspectives and at multiple levels of detail. Hominins could now put their own spin on the ideas of others, adapting them to individual needs and tastes, leading to cumulative innovation.

Thoughts, impressions, and attitudes could be modified by thinking about them in the context of each other, and they could be woven into an integrated "worldview" that defines who we are in relation to the world. This allowed the capacity for self expression, creating an environment conducive to the emergence of complex language, including grammar, recursion, word inflections, and syntactical structure, as well as comprehension. The proposal is consistent with findings that FOXP2 is associated with cognitive abilities that do not involve language, and with findings that non-language creative abilities arose at approximately the same time as complex language (Chrusch & Gabora, 2014). It is also consistent with findings that despite the existence of sophisticated cognitive abilities in other species such as birds (Emery, 2016), we alone exhibit cumulative cultural evolution. Cumulative cultural evolution may involve the 'recycling' of cortical maps such that cultural innovations invade evolutionarily older brain circuits and inherit some of their structural constraints (Dehaene, 2005; Lierberman, 2016). In short, we propose that the distinctive rich symbolism and grammatically complex language of the genus *Homo* reflect two evolutionary transitions brought about by novel forms of cognitive information processing.

Many evolutionary approaches to the general question of how modern cognition arose have been devised in the last few decades, such as those of Wynn and Coolidge (e.g., Wynn et al, 2017) highlighting developmental psychology) and Bruner (e.g. 2010), highlighting palaeoneurology; we submit our approach as one of this array of modern evolutionary approaches to the same broad issue of the origins of BM.

We presented archaeological evidence for the view that two cognitive transitions gave rise to two cultural transitions, as well as support for the proposed scenario obtained using an agent-based model of cultural evolution. Although such a model cannot provide proof it can play an important role in building a case by demonstrating the feasibility of the proposed mechanisms. Incorporating one kind of STR—chaining—into the computational model increased the fitness and diversity of cultural outputs, as well as the effectiveness of learning. The simulations, including the implementation of both chaining and CF, were simplistic; nevertheless the results suggest that once hominins became able to sequence thoughts together to generate increasingly complex and refined cultural outputs, and shift between different processing modes, they would have had a significant adaptive advantage. In future investigations we will use a sophisticated mathematical theory of concepts (Aerts, Gabora, & Sozzo, 2013) to develop a richer and more realistic model of cultural evolution. This will allow us to expand the simulation of STR to include not just chaining but also refinement of representations by viewing them from different contexts, and expand the simulation of the divergent mode of CF to incorporate the generation of new concept combinations.

We note that models of the origins of culture and BM have long suffered from vagueness. For example, Donald (1991) and Mithen (1996) both propose that hominin cognitive evolution arose in stages, but are vague as to how and why these stages unfolded. The transitions to possession of the cognitive capacities that we propose made BM possible—STR and CF—exhibit the defining characteristics of evolutionary transitions discussed in Section One, i.e., such transitions are rare, incomplete (did not 'throw a switch' resulting in immediate 'turning on' of BM), and involved new levels of organization. The increased sociality implied by the onset of STR and CF also meets Wilson's expectation that evolutionary transitions drive ". . . the suppression of fitness differences within groups, causing between-group selection to become the primary evolutionary force" (Wilson, 2010). It is interesting that EVOC results support Griesmer's (2000) hypothesis that a stage involving novel information complexity precedes stabilization mechanisms that 'fix' fit innovations as illustrated by the initial increase and subsequent decrease in cultural diversity.

It may well be that early models of the origins of hominin culture were accurate, but not precise, and that present-day precision reflects an emerging 'Extended Evolutionary Synthesis' (Smith & Ruppell, 2011; Woese, 2004). The origins of BM are currently being rethought in light of wide dissatisfaction with an archaic 'trait-list' approach to its understanding (Ames, Riel-Salvatore, & Collins, 2013) and with new, nonlinear models of multifaceted cultural evolutionary change (Mesoudi, 2009; McDowell, 2013). We propose that the origins of BM can be considered in terms of an evolutionary transition in which new varieties of information were generated and handled—both within the mind and in artificial memory systems external to it—to the degree that new social arrangements appeared.

Similarly, our theoretical arguments, and results obtained with EVOC, suggest that once humans became able to employ an exploratory, divergent processing mode when stuck, followed by a shift to a more constrained convergent processing mode to fine-tune their cultural outputs, they would have been capable of generating significantly more valuable cultural outputs. A potential pitfall of processing in a divergent mode is that since effort is devoted to the re-processing of previously learned material, less effort may be devoted to being on the lookout for danger and simply carrying out practical tasks. Since divergent thought carries a high cognitive load, it would not have been useful to think divergently until there was a means to shift back to a convergent mode. Although the EVOC results do not prove that onset of the capacity to chain thoughts together into sequences, and to shift between divergent modes of thought through CF, are responsible for our cultural complexity, it shows that they provide a computationally feasible explanation. We know of no other cognitive mechanisms implicated in the evolution of complex culture for which open-ended, adaptive cultural change has been demonstrated.

**Acknowledgments**

This work was supported by a grant (62R06523) to the first author from the Natural Sciences and Engineering Research Council of Canada.

## Appendix A:

## Training the Neural Network

The neural network starts with small random weights between input/output nodes. Weights between hidden nodes, and weights between hidden nodes and input/output nodes, are fixed at +/- 1.0. Patterns that represent ideas for actions are learned by training for 50 iterations using the generalized delta rule with a sigmoid activation function (Rumelhart & McClelland 1986). Since the network is an auto-associator training continues until the output matches the input. The relevant variables are:

$a_i$ = activation of $j$

$t_j$ = $j^{th}$ component of input

$w_{ij}$ = weight on link from $i$ to $j$

$\beta$ = 0.15

$\theta$ = 0.5

$$a_j = \frac{1}{(1 + e^{-\beta[\sum w_{ij} a_i + \theta]})} \tag{5}$$

For the movement node, we use the absolute value of $a_i$ (since negative movement is not possible; the least you can move is to not move at all). The comparison between input and output involves computing an error term, which is used to modify the pattern of connectivity in the network such that its responses become more correct. For input/output units the error term is computed as follows:

$$\delta_j = (t_j - a_j)a_j(1 - a_j) \tag{6}$$

For hidden units the error term is computed as follows:

$$\delta_i = a_j\,(1 - a_j)\sum \delta_j w_{ij} \tag{7}$$

## Appendix B:

## Calculating the Fitness of a Template

Assume that D is a sub-action (*i.e., $D = d_j$; $j$* = 1..6) and $T^i$ is the $i^{th}$ template (*i.e.,* $T^i = t^i{}_j$ ; $j$ = 1..6). Thus, $d_j$ represents the position of the $j^{th}$ body part and the value of $d_j$ can be either 0 (neutral), 1 (up), or -1 (down). Likewise, the value of $t^i{}_j$ can be 0, 1, -1, or * (unspecified). Accordingly, the fitness of sub-action D is obtained as per Eq. 8.

$$F(D) = \sum_{i=1}^{19} \Phi(T^i, D) \times \Omega(T^i) \tag{8}$$

As shown in this equation, fitness is a function of template weight ($\Phi(T^i, D)$) and template order ($\Omega(T^i)$).

### Template Weight

$\Phi(T^i, D)$ is a function that determines the weight of sub-action $D$ by comparing it with template $T^i$. This weight is set to one if each component of the sub-action (*i.e.,* $d_j$ ; $j$ = 1..6) either matches the corresponding component of the template (*i.e.,* $t^i{}_j$ ; $j$ = 1..6) or if the corresponding components of the template is unspecified (*i.e.,* $t^i{}_j = *$):

$$\Phi(T^i, D) = \begin{cases} 1 & if\ \forall t^i_j \in T^i : t^i_j = d_j\ or\ * \\ 0 & otherwise \end{cases} \tag{9}$$

### Template Order

$\Omega(T_i)$ computes the order of the template $T_i$ by counting the number of components that have a specified value (*i.e.,* $t_i \neq *$).

$$\Omega(T^i) = \sum_{j=1, t^i_j \neq *}^{6} t^i_j \tag{10}$$

The optimal sub-actions are {0, 1, −1, 1, −1, 1}, {0, 1, −1, 1, −1, −1}, {0, −1, 1, −1, 1, 1}, and {0, −1, 1, −1, 1, −1}.